\documentclass{article}
\usepackage{float}

\usepackage{PRIMEarxiv}

\usepackage[utf8]{inputenc} 
\usepackage[T1]{fontenc}    
\usepackage{amsmath,amssymb,amsthm}  
\usepackage{fullpage}       
\usepackage{graphicx}       
\usepackage{algorithm}      
\usepackage[noend]{algpseudocode} 
\usepackage{tikz}           
\usepackage[hidelinks]{hyperref} 
\usepackage{caption}        
\usepackage{subcaption}     
\usepackage{hyperref}    
\usepackage{orcidlink}

\title{Minimum Riesz s-Energy Subset Selection in Ordered Point Sets via Dynamic Programming}

\author{
  Michael T.M. Emmerich\orcidlink{0000-0002-7342-2090}\\
  Faculty of Information Technology, University of Jyväskylä, Finland\\
  \texttt{michael.t.m.emmerich@jyu.fi}\\
}
\date{\today}

\begin{document}
\maketitle

\begin{abstract}
We present a dynamic programming algorithm for selecting a representative subset of size 
\(k\) from a given set with \(n\) points such that the Riesz \(s\)-energy is near minimized. While NP-hard in general dimensions, the one-dimensional case can use the natural data ordering for efficient dynamic programming as an effective heuristic solution approach. This approach is then extended to problems related to two-dimensional Pareto front representations arising in biobjective optimization problems. Under the assumption of sorted (or non-dominated) input, the method typically yields near-optimal solutions in most cases. We also show that the approach avoids mistakes of greedy subset-selection by means of example. However, as we demonstrate, there are exceptions where DP does not identify the global minimum; for example, in one of our examples, the DP solution slightly deviates from the configuration found by a brute-force search.This is because the DP scheme's recurrence is approximate. The total time complexity of our algorithm is shown to be \(O(n^2 k)\). We provide computational examples with discontinuous Pareto fronts and an  open-source Python implementation, demonstrating the approximate DP algorithm's effectiveness across various problems with large point sets.
\bigskip

\end{abstract}

\section{Introduction}

Often it is desirable to select a representative subset of points from a larger set such that the points are as evenly spread as possible. A common measure to quantify this evenness is the \emph{Riesz \(s\)-energy} \cite{Hardin08,Falcon24}. For a set \(S\) of \(k\) points, the Riesz \(s\)-energy is defined as
\[
E(S)=\sum_{1\le p<q\le k}\frac{1}{d(P_p,P_q)^s},
\]
where \(d(P_p,P_q)\) denotes the distance between the points \(P_p\) and \(P_q\) and \(s>0\) is a parameter. In this paper we describe a dynamic programming (DP) approach to select a subset that (nearly) minimizes the Riesz \(s\)-energy. We begin with the one-dimensional (1D) case (points on the real line) and then extend the technique to two-dimensional (2D) Pareto fronts in the context of biobjective optimization.

It is worth noting that while the general problem is NP-hard \cite{Pereverdieva2024}, in the 1D case (or for suitably ordered 2D Pareto fronts) the structure of the data permits an efficient DP formulation. Moreover, our approach is motivated by promising results in related problems such as the maximum subset selection of the hypervolume indicator for non-dominated points in 2D \cite{Auger09}. Although the DP strategy yields optimal or near-optimal solutions in most cases, as we illustrate later, there exist exceptions where the DP solution deviates from the globally best configuration as determined by an exhaustive brute-force search.

The remainder of the paper is organized as follows. Section~2 presents the dynamic programming formulation and its proof of correctness for the one-dimensional case. Section~3 extends the approach to two-dimensional Pareto fronts, followed by detailed examples and computational analyses. Finally, Section~4 concludes with a summary of our contributions and a discussion of potential future work.

\section{Dynamic Programming for Subset Selection in 1D}

Dynamic programming (DP) is an optimization method that decomposes a complex problem into simpler, overlapping subproblems, solves each subproblem once, and stores the solutions for future use. This method is particularly effective for problems that exhibit \emph{optimal substructure}---that is, an optimal solution to the problem can be constructed from (near) optimal solutions to its subproblems---and \emph{overlapping subproblems}, where the same subproblems are solved repeatedly. In the context of our subset selection problem, the property that the best selection of \(r\) points ending at a given point can be built upon a (near) optimal selection of \(r-1\) points ending at an earlier point is key. This inherent structure allows us to design a DP scheme that efficiently (and in most cases optimally) minimizes the Riesz \(s\)-energy. For further details on the fundamentals of dynamic programming, see \cite{Bellman57}.

\subsection{Problem Statement}

Let 
\[
x_1 < x_2 < \cdots < x_n
\]
be a sorted set of points in \(\mathbb{R}\). For a given integer \(k\) (\(1 \le k < n\)) and a parameter \(s>0\), our goal is to choose a subset 
\[
S = \{ x_{i_1}, x_{i_2}, \dots, x_{i_k} \}, \quad \text{with } i_1 < i_2 < \cdots < i_k,
\]
that (nearly) minimizes
\[
E(S) = \sum_{1 \le p < q \le k} \frac{1}{\left(x_{i_q} - x_{i_p}\right)^s}.
\]
Because the energy increases sharply when points are close (due to the term \(1/(x_{i_q}-x_{i_p})^s\)), the objective is to select \(k\) points that are as spread out as possible.

\subsection{Dynamic Programming Algorithm}

We define the DP state as
\[
DP(i,r)=\text{(near) minimum energy achievable by selecting \(r\) points from } \{x_1, \dots, x_i\} \text{ with } x_i \text{ as the last point.}
\]

\paragraph{Base Case:} For \(r=1\), there are no pairs of points, hence 
\[
DP(i,1)=0 \quad \text{for } i=1,\dots,n.
\]

\paragraph{Recurrence:} For \(r\ge2\) and \(i\ge r\), assume that a (near) optimal \((r-1)\)-subset ending at \(x_p\) (with \(p < i\)) is known. When adding \(x_i\) as the \(r\)th point, the additional cost is
\[
\Delta(p,i)=\sum_{x \in S(p,r-1)} \frac{1}{(x_i - x)^s},
\]
where \(S(p,r-1)\) denotes the subset corresponding to \(DP(p, r-1)\). Hence, the recurrence is given approximatly by
\[
DP(i,r)\approx\min_{1 \le p < i} \left\{ DP(p,r-1) + \Delta(p,i) \right\}.
\]

\subsubsection*{Pseudocode}

\begin{algorithm}[H]
\caption{DP Subset Selection for Minimizing Riesz \(s\)-Energy in 1D}\label{alg:dp1d}
\begin{algorithmic}[1]
\Require Sorted points \(x_1,\dots,x_n\), integer \(k\), parameter \(s>0\).
\For{\(i=1\) to \(n\)}
    \State \(DP(i,1) \gets 0\)
    \State \(S(i,1) \gets \{ x_i \}\) \Comment{Record the subset for reconstruction.}
\EndFor
\For{\(r = 2\) to \(k\)}
    \For{\(i = r\) to \(n\)}
        \State \(DP(i,r) \gets +\infty\)
        \For{\(p = r-1\) to \(i-1\)}
            \State Compute \(\Delta(p,i) \gets \sum_{x\in S(p,r-1)} \frac{1}{(x_i - x)^s}\)
            \If{\(DP(p,r-1)+\Delta(p,i) < DP(i,r)\)}
                \State \(DP(i,r) \gets DP(p,r-1)+\Delta(p,i)\)
                \State \(S(i,r) \gets S(p,r-1) \cup \{ x_i \}\)
            \EndIf
        \EndFor
    \EndFor
\EndFor
\State \Return \(\min_{i=k,\dots,n} DP(i,k)\) and the corresponding subset.
\end{algorithmic}
\end{algorithm}

\subsection{Discussion on (Near) Optimality}

While the recurrence and the DP formulation would ideally guarantee that the selected subset is optimal with respect to the locally computed subproblems, the non-local interactions inherent in the Riesz \(s\)-energy may result in situations where the overall subset is only near optimal. In most cases, the DP approach yields the optimal subset, but there exist exceptions. For example, in the second TikZ illustration in Section~3, the DP algorithm selects a subset whose energy is near optimal; however, a brute-force search over the configuration space produces a different subset with slightly lower energy when we choose $5$ points, whereas for $3$ and $4$ points the results are correct. This observation underlines that the DP algorithm, due to its reliance on local decisions, may sometimes overlook the global optimum when non-local interactions are significant.

\subsection{Time Complexity}

The DP table is of size \(O(nk)\) and for each state \(DP(i,r)\) the algorithm examines up to \(O(n)\) previous states. With appropriate precomputation (or memoization) of the \(\Delta(p,i)\) values, the overall time complexity is \(O(n^2 k)\).

\subsection{Examples}

\subsubsection*{Example 1: \(x=[0,\,1,\,3,\,6]\), \(k=3\), \(s=1\)}

\paragraph{Brute-Force Calculation:}  
List all 3-point subsets and compute the energy:
\begin{itemize}
    \item \(\{0,1,3\}\):
    \[
    E=\frac{1}{1-0}+\frac{1}{3-0}+\frac{1}{3-1}=1+\frac{1}{3}+\frac{1}{2}\approx 1.8333.
    \]
    \item \(\{0,1,6\}\):
    \[
    E=1+\frac{1}{6-0}+\frac{1}{6-1}\approx 1.3667.
    \]
    \item \(\{0,3,6\}\):
    \[
    E=\frac{1}{3-0}+\frac{1}{6-0}+\frac{1}{6-3}\approx 0.8333.
    \]
    \item \(\{1,3,6\}\):
    \[
    E=\frac{1}{3-1}+\frac{1}{6-1}+\frac{1}{6-3}\approx 1.0333.
    \]
\end{itemize}
Thus, the brute-force analysis indicates that the subset \(\{0,3,6\}\) has the lowest energy, approximately \(0.8333\).

\paragraph{DP Computation:}
\begin{itemize}
    \item For \(r=1\): \(DP(i,1)=0\) for all \(i\).
    \item For \(r=2\):
    \begin{itemize}
        \item \(DP(2,2)=DP(1,1)+1/(1-0)=1\).
        \item \(DP(3,2)=\min\{DP(1,1)+1/(3-0),\,DP(2,1)+1/(3-1)\}=\min\{1/3,\,1/2\}\approx 0.3333\).
        \item \(DP(4,2)=\min\{DP(1,1)+1/(6-0),\,DP(2,1)+1/(6-1),\,DP(3,1)+1/(6-3)\}\approx \min\{0.1667,\,0.2,\,0.3333\}=0.1667\).
    \end{itemize}
    \item For \(r=3\):
    \begin{itemize}
        \item \(DP(3,3)=DP(2,2)+\left[1/(3-0)+1/(3-1)\right]\approx 1+ \left(1/3+1\right)\approx 1.8333\) (only possible with \(p=2\)).
        \item \(DP(4,3)=\min\{DP(2,2)+\Delta(2,4),\,DP(3,2)+\Delta(3,4)\}\).  
        
        With \(p=2\): \(\Delta(2,4)=1/(6-0)+1/(6-1)\approx 1/6+1/5\approx 0.3667\), total \(=1+0.3667=1.3667\).\\[1mm]
        With \(p=3\): \(\Delta(3,4)=1/(6-0)+1/(6-3)\approx 1/6+1/3\approx 0.5\), total \(=0.3333+0.5=0.8333\).  
    \end{itemize}
\end{itemize}
Thus, the DP method finds an energy of \(0.8333\) for \(k=3\), confirming the brute-force result in this case.

\subsubsection*{Example 2: \(x=[0,\,2,\,4,\,7]\), \(k=3\), \(s=1\)}

\paragraph{Brute-Force Calculation:}
\begin{itemize}
    \item \(\{0,2,4\}\):
    \[
    E=\frac{1}{2-0}+\frac{1}{4-0}+\frac{1}{4-2}=\frac{1}{2}+\frac{1}{4}+\frac{1}{2}=1.25.
    \]
    \item \(\{0,2,7\}\):
    \[
    E=\frac{1}{2-0}+\frac{1}{7-0}+\frac{1}{7-2}\approx \frac{1}{2}+\frac{1}{7}+\frac{1}{5}\approx 0.8429.
    \]
    \item \(\{0,4,7\}\):
    \[
    E=\frac{1}{4-0}+\frac{1}{7-0}+\frac{1}{7-4}\approx \frac{1}{4}+\frac{1}{7}+\frac{1}{3}\approx 0.7262.
    \]
    \item \(\{2,4,7\}\):
    \[
    E=\frac{1}{4-2}+\frac{1}{7-2}+\frac{1}{7-4}\approx \frac{1}{2}+\frac{1}{5}+\frac{1}{3}\approx 1.0333.
    \]
\end{itemize}
The brute-force search indicates that the subset \(\{0,4,7\}\) is best with an energy of approximately \(0.7262\).

\paragraph{DP Computation:}
\begin{itemize}
    \item For \(r=1\): \(DP(i,1)=0\).
    \item For \(r=2\):
    \begin{itemize}
        \item \(DP(2,2)=0+1/(2-0)=0.5\).
        \item \(DP(3,2)=\min\{0+1/(4-0),\,0+1/(4-2)\}=\min\{0.25,\,0.5\}=0.25\).
        \item \(DP(4,2)=\min\{0+1/(7-0),\,0+1/(7-2),\,0+1/(7-4)\}=\min\{0.1429,\,0.2,\,0.3333\}=0.1429\).
    \end{itemize}
    \item For \(r=3\):
    \begin{itemize}
        \item \(DP(3,3)=DP(2,2)+\Delta(2,3)=0.5+\left[1/(4-0)+1/(4-2)\right]=0.5+0.75=1.25\).
        \item \(DP(4,3)=\min\{DP(2,2)+\Delta(2,4),\,DP(3,2)+\Delta(3,4)\}\).\\[1mm]
        With \(p=2\): \(\Delta(2,4)=1/(7-0)+1/(7-2)=0.1429+0.2=0.3429\), total \(=0.5+0.3429=0.8429\).\\[1mm]
        With \(p=3\): \(\Delta(3,4)=1/(7-0)+1/(7-4)=0.1429+0.3333=0.4762\), total \(=0.25+0.4762=0.7262\).
    \end{itemize}
\end{itemize}
In this case, the DP method finds a subset with energy \(0.7262\), matching the brute-force result. Note, however, that in other scenarios (as illustrated later) the DP approach may yield a near-optimal solution that slightly deviates from the brute-force optimum.

\section{Dynamic Programming for Subset Selection on 2D Pareto Fronts}

\subsection{Introduction}

In many biobjective optimization problems, the goal is to approximate the Pareto front --- a set of non-dominated solutions --- with a representative subset. Although the Pareto front is two-dimensional, when the points are sorted by one coordinate (typically \(f_1\)) the other coordinate (\(f_2\)) follows a monotonic order. This observation allows us to extend the 1D DP scheme to the 2D case.

\subsection{Problem Statement}

Let
\[
P_1, P_2, \dots, P_n \in \mathbb{R}^2
\]
be a set of non-dominated points, sorted such that 
\[
f_1(P_1) \le f_1(P_2) \le \cdots \le f_1(P_n)
\]
and, because of non-domination, 
\[
f_2(P_1) \ge f_2(P_2) \ge \cdots \ge f_2(P_n).
\]
For a given integer \(k\) (\(1 \le k < n\)) and a parameter \(s>0\), we wish to select a subset
\[
S=\{P_{i_1},P_{i_2},\dots,P_{i_k}\}, \quad i_1 < i_2 < \cdots < i_k,
\]
that (nearly) minimizes
\[
E(S)=\sum_{1\le p<q\le k}\frac{1}{d(P_{i_p},P_{i_q})^s},
\]
where \(d(P,Q)\) denotes the Euclidean distance between points \(P\) and \(Q\).

\subsection{Dynamic Programming Algorithm}

Similarly, we define the DP state by
\[
DP(i,r)=\text{(near) minimum energy attainable by selecting \(r\) points from } \{P_1,\dots,P_i\} \text{ with } P_i \text{ as the last point.}
\]
We also record the corresponding subset \(S(i,r)\).

\paragraph{Base Case:} For \(r=1\),
\[
DP(i,1)=0 \quad \text{for } i=1,\dots,n,
\]
with \(S(i,1)=\{P_i\}\).

\paragraph{Recurrence:} For \(r\ge2\) and \(i\ge r\), assume a (near) optimal \((r-1)\)-subset ending at \(P_p\) (with \(p < i\)) is computed. When adding \(P_i\), the incremental cost is given by
\[
\Delta(p,i)=\sum_{P\in S(p,r-1)} \frac{1}{d(P,P_i)^s}.
\]
Thus, the recurrence becomes (due to non-local interactions we can only state this approximately):
\[
DP(i,r)\approx\min_{p=r-1,\ldots,i-1}\{DP(p,r-1)+\Delta(p,i)\},
\]
with
\[
S(i,r)=S(p,r-1)\cup\{P_i\}.
\]

\subsubsection*{Pseudocode}

\begin{algorithm}[H]
\caption{DP Subset Selection for Minimizing Riesz \(s\)-Energy on a 2D Pareto Front}\label{alg:dp2d}
\begin{algorithmic}[1]
\Require Non-dominated points \(P_1,\dots,P_n \in \mathbb{R}^2\) sorted such that 
\[
f_1(P_1) \le \cdots \le f_1(P_n) \quad\text{and}\quad f_2(P_1) \ge \cdots \ge f_2(P_n),
\]
integer \(k\), parameter \(s>0\).
\For{\(i=1\) to \(n\)}
    \State \(DP(i,1) \gets 0\)
    \State \(S(i,1) \gets \{P_i\}\)
\EndFor
\For{\(r = 2\) to \(k\)}
    \For{\(i = r\) to \(n\)}
        \State \(DP(i,r) \gets +\infty\)
        \For{\(p = r-1\) to \(i-1\)}
            \State Compute \(\Delta(p,i) \gets \displaystyle \sum_{P\in S(p,r-1)} \frac{1}{d(P,P_i)^s}\)
            \If{\(DP(p,r-1)+\Delta(p,i) < DP(i,r)\)}
                \State \(DP(i,r) \gets DP(p,r-1)+\Delta(p,i)\)
                \State \(S(i,r) \gets S(p,r-1)\cup\{P_i\}\)
            \EndIf
        \EndFor
    \EndFor
\EndFor
\State \Return \(\min_{i=k,\dots,n}\{DP(i,k)\}\) and the corresponding subset.
\end{algorithmic}
\end{algorithm}

\subsection{Discussion on (Near) Optimality in 2D}

Under the assumption of sorted (or non-dominated) input, the DP algorithm constructs the solution by combining locally optimal decisions. In many cases, this leads to an overall near-optimal or even optimal subset. However, due to the non-local interactions present in the Riesz \(s\)-energy, there are instances where the globally best configuration (as determined by a brute-force search) differs from the DP-selected subset. In our experiments, most candidate subsets are near-optimal; nevertheless, as illustrated in the second TikZ figure (Fig.~\ref{fig:subset3}), a brute-force approach can yield a different configuration with marginally lower energy.

\subsection{Heuristic Justification for Approximate Optimality}

Our approach is designed to efficiently construct a subset \(S\) of size \(k\) that approximately minimizes the energy
\[
E(S) = \sum_{1 \le p < q \le k} \frac{1}{d(P_{i_p}, P_{i_q})^s},
\]
where the points \(P_1, \dots, P_n \in \mathbb{R}^2\) are non-dominated and sorted by increasing \(f_1\) (and, consequently, by decreasing \(f_2\)). This ordering guarantees that for any indices \(i < j\), we have
\[
f_1(P_i) \le f_1(P_j) \quad \text{and} \quad f_2(P_i) \ge f_2(P_j).
\]
In particular, this structure yields a natural lower bound on the distances between points:
\[
d(P_i, P_j) \ge f_1(P_j) - f_1(P_i).
\]

This property is central to the effectiveness of our dynamic programming scheme. When extending a subset ending at a point \(P_p\) (with \(p < i\)) to include a new point \(P_i\), the additional cost—denoted by \(\Delta(p,i)\)—tends to behave in an approximately monotonic manner with respect to the ordering. Consequently, a (near) optimal subset ending at \(P_i\) can be obtained by selecting the best extension from some (near) optimal subset ending at an earlier point.

The resulting recurrence,
\[
DP(i, r) \approx \min_{p < i} \{ DP(p, r-1) + \Delta(p,i) \},
\]
captures this intuition. While non-local interactions can cause minor deviations from the global energy minimum, the method typically produces energy values close to or often equal to the optimum.

\subsection{Example: Biobjective Optimization}

Consider the following set of non-dominated points:
\[
P_1=(1,15),\quad P_2=(5,10),\quad P_3=(8,4),\quad P_4=(13,3),\quad P_5=(15,2),\quad P_6=(17,1).
\]
They are sorted by the first coordinate. Suppose we wish to select \(k=3\) points. One candidate selected by the DP approach is 
\[
S=\{P_1,P_3,P_6\}=\{(1,15),(8,4),(17,1)\}.
\]
With approximate pairwise distances:
\[
d(P_1,P_3)\approx 13.04,\quad d(P_1,P_6)\approx 21.26,\quad d(P_3,P_6)\approx 9.49,
\]
the energy (with \(s=1\)) is approximately
\[
E(S)\approx \frac{1}{13.04}+\frac{1}{21.26}+\frac{1}{9.49}\approx 0.229.
\]

\begin{figure}[h]
  \centering
  \begin{minipage}{0.5\textwidth}
    \centering
    \begin{tikzpicture}[scale=0.38, every node/.style={font=\large}]
      \draw[dotted, step=1, gray] (0,0) grid (20,20);
      \draw[->, thick] (0,0) -- (20,0) node[right] {\(f_1\)};
      \draw[->, thick] (0,0) -- (0,20) node[above] {\(f_2\)};
      
      \foreach \x/\y/\name in {1/15/\(P_1\), 5/10/\(P_2\), 8/4/\(P_3\), 13/3/\(P_4\), 15/2/\(P_5\), 17/1/\(P_6\)}
      {
        \draw[black, fill=black] (\x,\y) circle (6pt);
        \node[above right] at (\x,\y) {\name};
      }
      
      \foreach \x/\y in {1/15, 8/4, 17/1}
      {
        \draw[red, thick] (\x,\y) circle (8pt);
      }
      
      \node[red] at (10,16) {DP Subset: \(\{P_1,P_3,P_6\}\)};
    \end{tikzpicture}
  \end{minipage}%
  \hfill
  \begin{minipage}{0.5\textwidth}
    \centering
    \begin{tikzpicture}[scale=0.38, every node/.style={font=\large}]
      \draw[dotted, step=1, gray] (0,0) grid (20,20);
      \draw[->, thick] (0,0) -- (20,0) node[right] {\(f_1\)};
      \draw[->, thick] (0,0) -- (0,20) node[above] {\(f_2\)};
      
      \draw[black, fill=black] (2,20) circle (6pt) node[above left] {\(P_1\)};
      \draw[black, fill=black] (4,18) circle (6pt) node[above left] {\(P_2\)};
      \draw[black, fill=black] (6,16) circle (6pt) node[above left] {\(P_3\)};
      \draw[black, fill=black] (9,12) circle (6pt) node[above left] {\(P_4\)};
      \draw[black, fill=black] (11,8) circle (6pt) node[below right] {\(P_5\)};
      \draw[black, fill=black] (14,5) circle (6pt) node[below right] {\(P_6\)};
      \draw[black, fill=black] (17,3) circle (6pt) node[below right] {\(P_7\)};
      
      \draw[red, thick] (2,20) circle (8pt);
      \draw[red, thick] (9,12) circle (8pt);
      \draw[red, thick] (17,3) circle (8pt);
      
    \end{tikzpicture}
  \end{minipage}
  \caption{\label{fig:subset3}DP-selected representative subsets on two Pareto front examples (with \(k=3\)). The right illustration shows an instance where the DP method yields a near-optimal solution that deviates from the brute-force result.}

  \centering
  \begin{minipage}{0.5\textwidth}
    \centering
    \begin{tikzpicture}[scale=0.38, every node/.style={font=\large}]
      \draw[dotted, step=1, gray] (0,0) grid (20,20);
      \draw[->, thick] (0,0) -- (20,0) node[right] {\(f_1\)};
      \draw[->, thick] (0,0) -- (0,20) node[above] {\(f_2\)};
      \draw[black, fill=black] (1,15) circle (6pt) node[above right] {\(P_1\)};
      \draw[black, fill=black] (5,10) circle (6pt) node[above right] {\(P_2\)};
      \draw[black, fill=black] (8,4)  circle (6pt) node[above right] {\(P_3\)};
      \draw[black, fill=black] (13,3) circle (6pt) node[above right] {\(P_4\)};
      \draw[black, fill=black] (15,2) circle (6pt) node[above right] {\(P_5\)};
      \draw[black, fill=black] (17,1) circle (6pt) node[above right] {\(P_6\)};
      \draw[red, thick] (1,15) circle (8pt);
      \draw[red, thick] (5,10) circle (8pt);
      \draw[red, thick] (8,4)  circle (8pt);
      \draw[red, thick] (17,1) circle (8pt);
    \end{tikzpicture}
  \end{minipage}%
  \hfill
  \begin{minipage}{0.5\textwidth}
    \centering
    \begin{tikzpicture}[scale=0.38, every node/.style={font=\large}]
      \draw[dotted, step=1, gray] (0,0) grid (20,20);
      \draw[->, thick] (0,0) -- (20,0) node[right] {\(f_1\)};
      \draw[->, thick] (0,0) -- (0,20) node[above] {\(f_2\)};
      \draw[black, fill=black] (2,20) circle (6pt) node[above left] {\(P_1\)};
      \draw[black, fill=black] (4,18) circle (6pt) node[above left] {\(P_2\)};
      \draw[black, fill=black] (6,16) circle (6pt) node[above left] {\(P_3\)};
      \draw[black, fill=black] (9,12) circle (6pt) node[above left] {\(P_4\)};
      \draw[black, fill=black] (11,8) circle (6pt) node[above left] {\(P_5\)};
      \draw[black, fill=black] (14,5) circle (6pt) node[above left] {\(P_6\)};
      \draw[black, fill=black] (17,3) circle (6pt) node[above left] {\(P_7\)};
      \draw[red, thick] (2,20) circle (8pt);
      \draw[red, thick] (6,16) circle (8pt);
      \draw[red, thick] (11,8) circle (8pt);
      \draw[red, thick] (17,3) circle (8pt);
    \end{tikzpicture}
  \end{minipage}
  \caption{\label{fig:subset4}DP-selected representative subsets on two Pareto front examples for \(k=4\).}
  \bigskip
\end{figure}
\newpage
In the second example we consider the set $P = {(2,20), (4,18), (6,16), (9,12), (11,8), (14,5), (17,3)}$.

The results are depicted in Fig.~\ref{fig:subset3}. Moreover, selected point sets for \(k=4\) are shown in Fig.~\ref{fig:subset4}. Interestingly, the set of points selected for \(k=3\) differs from the one for \(k=4\). These examples also illustrate that employing a greedy algorithm will not work. Note that all examples have been verified with brute-force computations; in all depicted cases the DP-selected subsets are  optimal.

\subsection{Counterexample}
\begin{figure}[t]
\centering
\begin{minipage}{0.48\textwidth}
    \centering
    \textbf{DP Solution}\\[3mm]
    \begin{tikzpicture}[scale=0.38, every node/.style={font=\large}]
      \draw[dotted, step=1, gray] (0,0) grid (20,20);
      \draw[->, thick] (0,0) -- (20,0) node[right] {\(f_1\)};
      \draw[->, thick] (0,0) -- (0,20) node[above] {\(f_2\)};
      
      \draw[black, fill=black] (2,20) circle (6pt) node[above left] {\(P_1\)};
      \draw[black, fill=black] (4,18) circle (6pt) node[above left] {\(P_2\)};
      \draw[black, fill=black] (6,16) circle (6pt) node[above left] {\(P_3\)};
      \draw[black, fill=black] (9,12) circle (6pt) node[above left] {\(P_4\)};
      \draw[black, fill=black] (11,8) circle (6pt) node[above left] {\(P_5\)};
      \draw[black, fill=black] (14,5) circle (6pt) node[above left] {\(P_6\)};
      \draw[black, fill=black] (17,3) circle (6pt) node[above left] {\(P_7\)};
      
      \draw[red, thick] (2,20) circle (8pt);
      \draw[red, thick] (6,16) circle (8pt);
      \draw[red, thick] (9,12) circle (8pt);
      \draw[red, thick] (11,8) circle (8pt);
      \draw[red, thick] (17,3) circle (8pt);
    \end{tikzpicture}
\end{minipage}
\hfill
\begin{minipage}{0.48\textwidth}
    \centering
    \textbf{Brute Force Solution}\\[3mm]
    \begin{tikzpicture}[scale=0.38, every node/.style={font=\large}]
      \draw[dotted, step=1, gray] (0,0) grid (20,20);
      \draw[->, thick] (0,0) -- (20,0) node[right] {\(f_1\)};
      \draw[->, thick] (0,0) -- (0,20) node[above] {\(f_2\)};
      
      \draw[black, fill=black] (2,20) circle (6pt) node[above left] {\(P_1\)};
      \draw[black, fill=black] (4,18) circle (6pt) node[above left] {\(P_2\)};
      \draw[black, fill=black] (6,16) circle (6pt) node[above left] {\(P_3\)};
      \draw[black, fill=black] (9,12) circle (6pt) node[above left] {\(P_4\)};
      \draw[black, fill=black] (11,8) circle (6pt) node[above left] {\(P_5\)};
      \draw[black, fill=black] (14,5) circle (6pt) node[above left] {\(P_6\)};
      \draw[black, fill=black] (17,3) circle (6pt) node[above left] {\(P_7\)};
      
      \draw[red, thick] (2,20) circle (8pt);
      \draw[red, thick] (6,16) circle (8pt);
      \draw[red, thick] (9,12) circle (8pt);
      \draw[red, thick] (14,5) circle (8pt);
      \draw[red, thick] (17,3) circle (8pt);
    \end{tikzpicture}
\end{minipage}
\caption{\label{fig:counterex}Comparison of the DP and Brute Force solutions for Example 2. In the DP solution, points \(P_1\), \(P_3\), \(P_4\), \(P_5\), and \(P_7\) are selected, while in the Brute Force solution, points \(P_1\), \(P_3\), \(P_4\), \(P_6\), and \(P_7\) are selected.}
\end{figure}
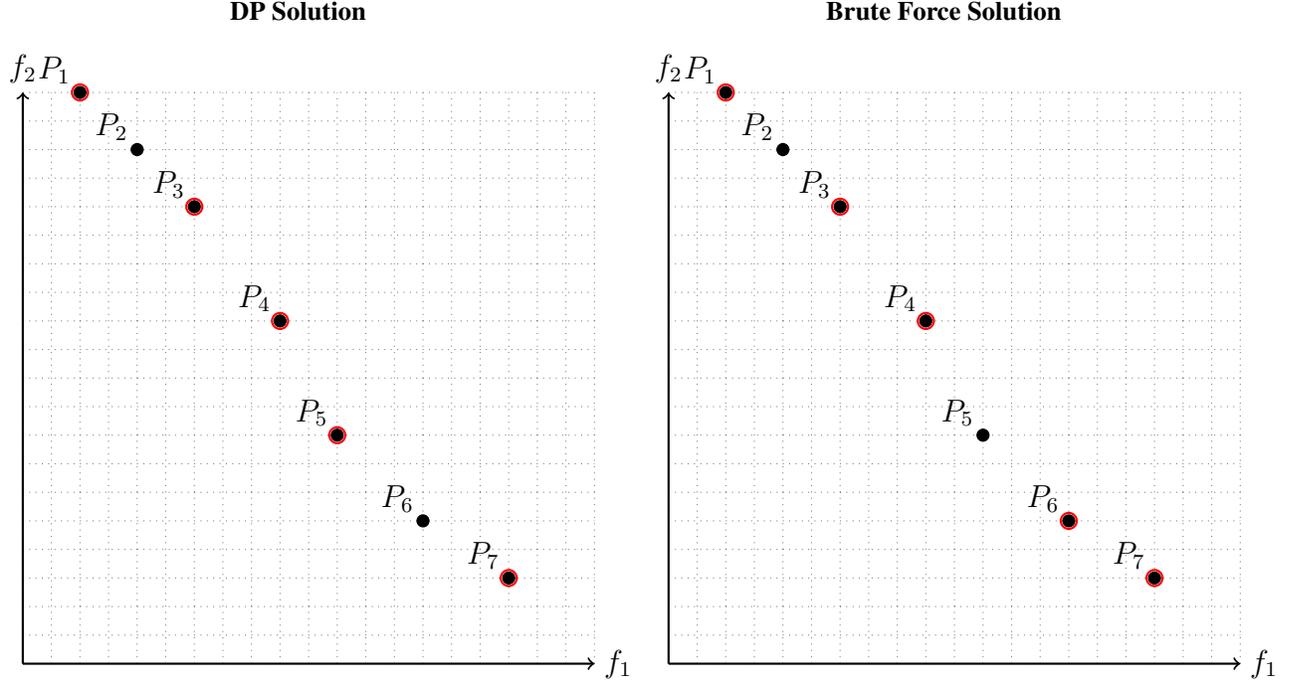

Next we show that the idea of using DP for subset selection does not always produce the optimal results, unlike in problems where the recursion is precise and not approximate (e.g., in the 2-D hypervolume-indicator subset selection  \cite{Auger09}).
We note that it is not easy to find a counterexample, and that the error of the DP algorithm is very small.
Consider the set $P = {(2,20), (4,18), (6,16), (9,12), (11,8), (14,5), (17,3)}$ that was used in the second example. Now we chose $k=5$ out of the $n=7$ points, and this will yield different results. See \ref{fig:counterex} for a visualization and the details of the computation below:
\noindent\begin{minipage}{\textwidth}
\textbf{Dynamic Programming (DP) Solution:}
\begin{itemize}
    \item \emph{Selected subset indices:} [0, 2, 3, 4, 6]
    \item \emph{Selected subset points:} $\{(2, 20), (6, 16), (9, 12), (11, 8), (17, 3)\}$
    \item \emph{Total Riesz $s$-energy (with $s=1$):} 1.181
\end{itemize}

\vspace{1em}

\noindent\textbf{Brute Force Solution:}
\begin{itemize}
    \item \emph{Selected subset indices:} [0, 2, 3, 5, 6]
    \item \emph{Selected subset points:} $\{(2, 20), (6, 16), (9, 12), (14, 5), (17, 3)\}$
    \item \emph{Total Riesz $s$-energy (with $s=1$):} 1.176
\end{itemize}
\end{minipage}

\noindent Since the DP solution does \textbf{not} match the brute-force solution, this example serves as a counterexample for the DP finding always the minimum subset.

\subsection{Example with larger dataset}
Next, we demonstrate the effectiveness of the algorithm in computing a sparse, representative, and evenly distributed subset from a larger set of points underpinning its practical value.

\begin{figure}[htbp]
  \centering
  \includegraphics[width=0.8\textwidth]{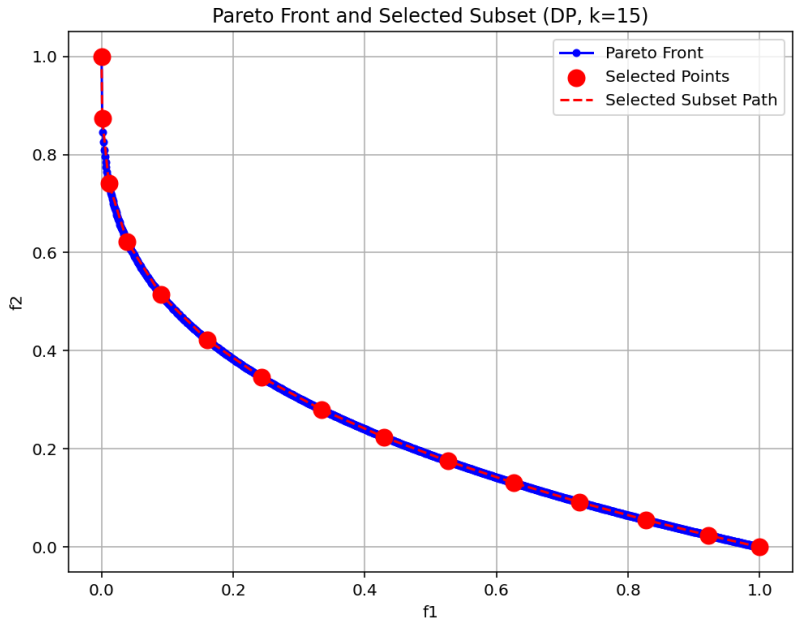}
  \caption{Pareto front subset selection for a set of 1000 points (blue). The red points have been selected by the DP algorithm.}
  \label{fig:pareto_subset}
\end{figure}

The Pareto front approximation in Figure \ref{fig:pareto_subset} was generated using 1000 points from the function 
  \[
    f_2 = 1 - f_1^{0.3}, \quad f_1 \in [0,1],
  \]
  consisting of mutually non-dominated points.
  The blue points represent the full Pareto front approximation set. A dynamic programming algorithm selects 15 representative points (highlighted in red) that minimize the Riesz $s$-energy (with $s=1$). DP is selecting a subset, see Figure \ref{fig:pareto_subset}. The $15$ selected points are further connected by a dashed red line to illustrate the representative subset path.
This example shows that the DP algorithm is very effective in selecting a well-separated subset of the Pareto front approximation even for larger point sets.

A second example shall investigate the behavior of the DP algorithm on a discontinuous Pareto front approximation.

\subsection{ZDT3 Pareto Front and DP-Selected Representative Subset}

The ZDT3 problem \cite{Zitzler00} is a well-known benchmark in multiobjective optimization, characterized by a discontinuous Pareto front. The Pareto front is defined by
\begin{equation}
\begin{aligned}
f_1(x) &= x_1, \\
f_2(x) &= 1 - \sqrt{x_1} - x_1 \sin(10\pi x_1),
\end{aligned}
\end{equation}
with the decision variable \(x_1\) restricted to the following intervals:
\begin{equation}
\begin{aligned}
I_1 &= [0.0,\;0.0830015349], \\
I_2 &= [0.1822287280,\;0.2577623634], \\
I_3 &= [0.4093136748,\;0.4538821041], \\
I_4 &= [0.6183967944,\;0.6525117038], \\
I_5 &= [0.8233317983,\;0.8518328654].
\end{aligned}
\end{equation}

To study this front, we generated an approximation consisting of 1,000 points sampled evenly across these intervals. A dynamic programming algorithm that minimizes the Riesz \(s\)-energy (with \(s=1\)) was then applied to select 15 representative points which effectively capture the structure of the front.

\begin{figure}[ht!]
    \centering
    \includegraphics[width=0.8\textwidth]{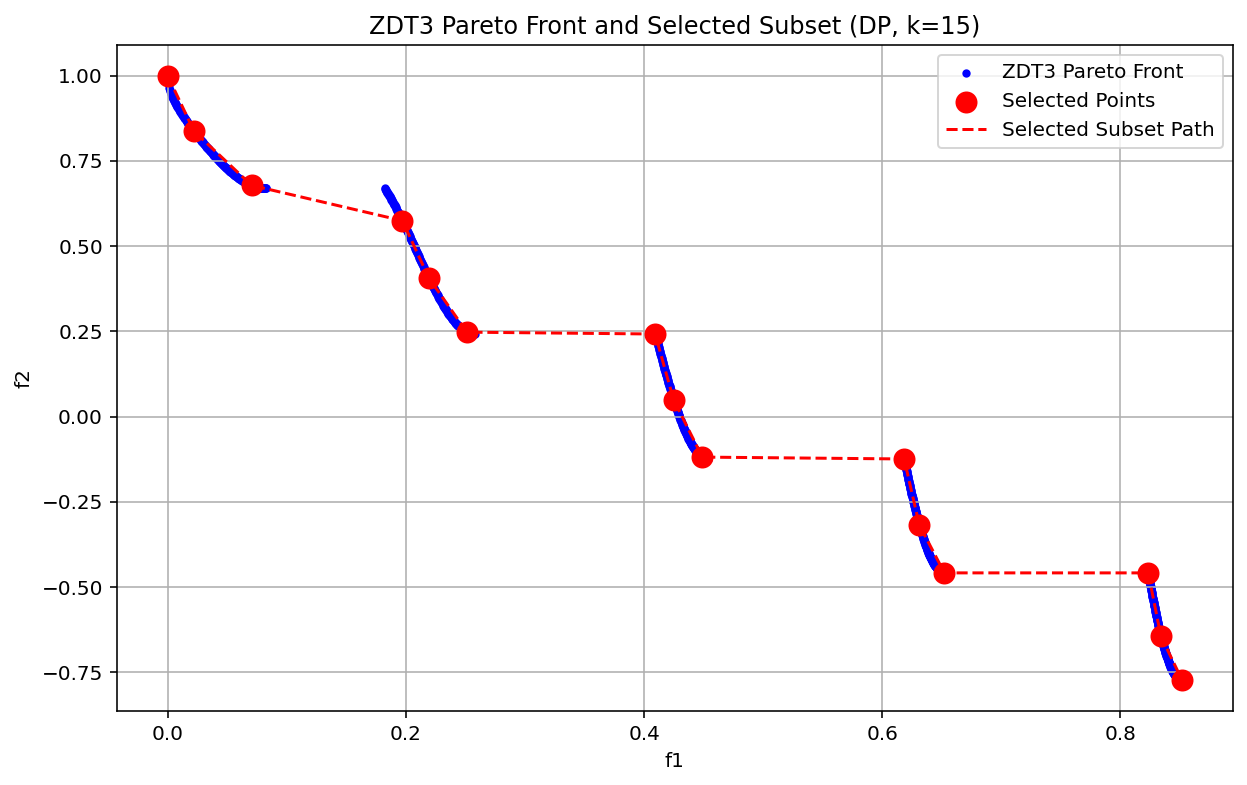} 
    \caption{ZDT3 Pareto Front and DP-Selected Representative Subset. The blue scatter plot displays 1,000 points approximating the ZDT3 Pareto front, defined by \(f_1\) over five disjoint intervals and \(f_2 = 1 - \sqrt{f_1} - f_1\sin(10\pi f_1)\). The red markers represent the 15 representative points selected by a dynamic programming algorithm that minimizes the Riesz 1-energy, with a dashed line connecting them in the order they were chosen.}
    \label{fig:zdt3}
\end{figure}

Figure~\ref{fig:zdt3} illustrates the full approximation of the ZDT3 Pareto front (blue points) along with the DP-selected representative subset (red markers). The red markers are connected by a dashed line indicating their sequential ordering. Notably, the points capture all segments, showing the practical value of our DP approach. However, due to the combinatorial nature of the problem, we could not double check with the brute-force algorithm to see how much the error deviates from the optimum result.

\begin{figure}[ht!]
    \centering
    \includegraphics[width=0.8\textwidth]{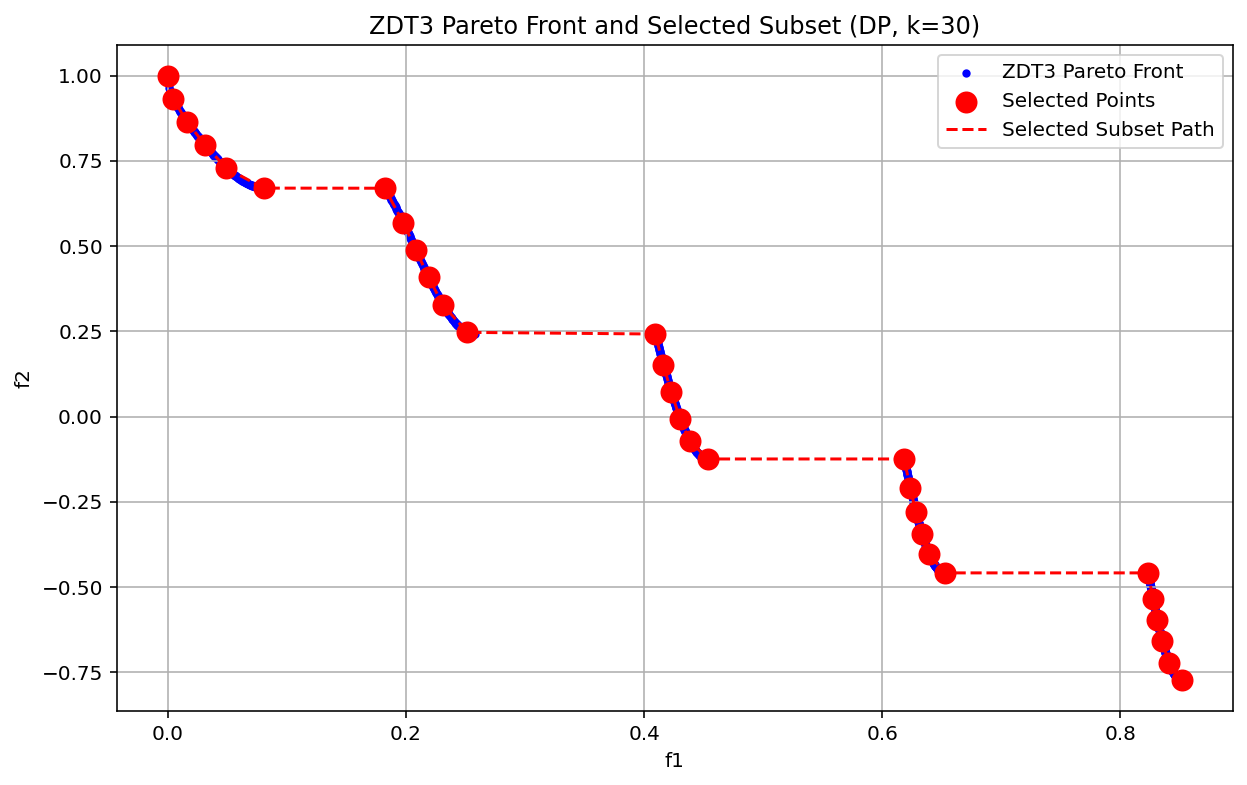} 
    \caption{ZDT3 Pareto Front and DP-Selected Representative Subset (k=30).}
    \label{fig:zdt3_k30}
\end{figure}

In addition, we conducted an experiment with a larger subset size of \(k=30\) representative points while still sampling 1,000 points from the Pareto front. This larger subset provides a more detailed representation of the discontinuous ZDT3 front, capturing additional nuances and variations in the objective space.

Figure~\ref{fig:zdt3_k30} shows the full approximation of the ZDT3 Pareto front (blue points) along with the DP-selected subset of 30 representative points (red markers). The red markers are connected by a dashed line to illustrate the sequential ordering of the selected points, thereby highlighting the improved granularity in representing the Pareto front.

\section{Conclusion}

We presented a dynamic programming algorithm for subset selection that nearly minimizes the Riesz \(s\)-energy on 1D and 2D Pareto fronts. In 1D, data ordering permits a simple DP approach, while in 2D, sorting by increasing \(f_1\) and decreasing \(f_2\) enables a similar strategy. In most cases the method yields near-optimal, and often optimal, solutions; also the solutions seem to be better than those obtained by greedy subset selection, as our examples show. However, due to the non-local interactions inherent in the Riesz \(s\)-energy, there are cases where the DP-selected subset deviates from the globally best configuration (as verified by brute-force search, see e.g. Fig.~\ref{fig:subset3}). Future work may explore hybrid methods or enhanced DP recurrences that better capture global interactions and further improve the quality of the solution.
However, our paper shows that DP, even if the recurrence is not exact, can lead to effective approximations.

It is an open question whether the problem can be solved faster or whether it allows for an exact polynomial-time algorithm. One may, for instance, consider using mechanisms such as in \cite{Kuhn16,Bringmann14} or linear programming. Secondly, it is not known whether there is a 'nerve' or 'wire' structure \cite{Wegner75,Bringmann17} in more general problems, which would allow one to build logical circuits and thus establish the NP hardness in geometrical settings for one or more dimensions.

\subsection{Data Sets}
A permanent link to the \texttt{Python} implementation of all algorithms and examples, including the brute-force algorithm (for verification purposes), is available \cite{Emmerich25}.

\bibliographystyle{plain}

\end{document}